\documentclass[aps,prl,reprint,superscriptaddress,nofootinbib]{revtex4-1}
\usepackage{amsmath}
\usepackage{graphicx}
\usepackage{units}  
\usepackage{xspace}
\usepackage{subfigure}
\usepackage{hyperref}
\newcommand{\mb}{\mathbf}

\newcommand{\beq}{\begin{equation}}
\newcommand{\eeq}{\end{equation}}
\newcommand{\bea}{\begin{eqnarray}}
\newcommand{\eea}{\end{eqnarray}}

\usepackage[usenames]{color}

\begin{document}
\bibliographystyle{apsrev}
 
\title{Emergent geometric frustration and flat band in moir{\'e} bilayer graphene}

\author{Hridis K. Pal}
\affiliation{School of Physics, Georgia Institute of Technology, Atlanta, GA 30332-0430, USA}
\affiliation{Department of Physics, University of Houston, Houston, Texas 77204, USA}
\email{hridis.pal@physics.gatech.edu}
\email{markus.kindermann@physics.gatech.edu}
\author{Stephen Spitz}
\affiliation{School of Physics, Georgia Institute of Technology, Atlanta, GA 30332-0430, USA}
\author{Markus Kindermann}
\affiliation{School of Physics, Georgia Institute of Technology, Atlanta, GA 30332-0430, USA}

\begin{abstract}
So far the physics of moir{\'e} graphene bilayers at large, incommensurate rotation angles has been considered uninteresting. It has been held that the interlayer coupling in such structures is weak and the system can be thought of as a pair of decoupled single graphene sheets to a good approximation.  Here, we demonstrate that for large rotation angles near commensurate ones,  the interlayer coupling, far from being weak, is able to completely localize electrons to within a large scale, geometrically frustrated network of topologically protected modes. The  emergent geometric frustration of the system gives rise to completely flat bands, with  strong correlation physics as a result. All of this arises although in the lattice structure no large scale  pattern appears  to the unguided eye. Sufficiently close to commensuration the low-energy physics of this remarkable system has an exact analytical solution.
\end{abstract}

\maketitle 

When two graphene layers are placed on top of each other and rotated relative to one another, they exhibit beautiful moir{\'e} patterns. Moir{\'e} bilayers fall into one of two categories: commensurate, where lattice periodicity is present, and incommensurate, where lattice periodicity is absent \cite{lopes, shallcross, meleprb, laissardiere, bistritzer}. At commensurate angles, the low-energy physics depends critically on the sublattice exchange symmetry (SE) of the structure \cite{meleprb}: SE even (SEE) structures are gapped  \cite{meleprb} and are topological in nature \cite{kindermann:prl15}, while SE odd (SEO)  structures are ungapped. In incommensurate bilayers, on the other hand, it is generally held that the effect of the interlayer coupling is weak at large angles but increases with decreasing twist angle, i.e., with increasing moir{\'e} size. Indeed, it has been shown \cite{lopes,luican}   that  as the rotation angle decreases, the interlayer motion  increasingly suppresses the charge carrier velocity. When the angle becomes of order   $1^{\circ}$, the electron velocity vanishes altogether at certain magic angles \cite{bistritzer}, leading to localization of electrons \cite{laissardiere, morell, bistritzer}. Recent  experiments have observed such localization with strong correlations as a result:   a Mott state \cite{mott} and an unconventional superconducting state \cite{supercon}. This has given added impetus to this exciting field.

\begin{figure}
\begin{center}
 \includegraphics[scale=0.36,trim={0 6.cm 10cm 0},clip]{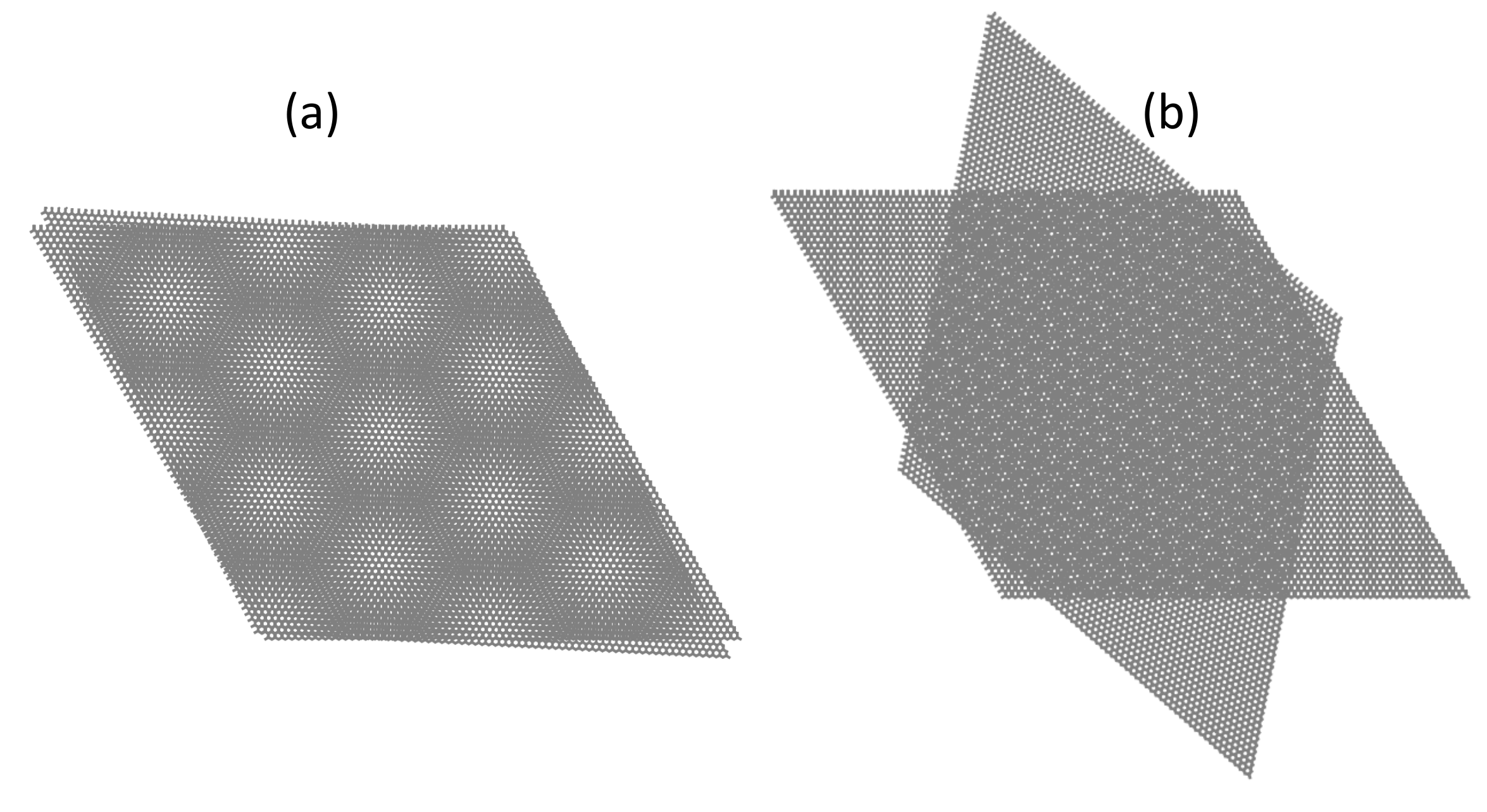}
\caption{Moir{\'e} patterns obtained by rotating two graphene layers by $3^{\circ}$ away from commensuration angles (a) $\theta_c=0^{\circ}$ and (b) $\theta_c=38.21^{\circ}$. While near $\theta_c=0^{\circ}$, a large-scale pattern appears, periodically repeating AA- and AB-like regions, the lattice near $\theta_c=38.21^{\circ}$  is visually featureless. In the text we demonstrate that nevertheless, as one approaches closer to $\theta_c$, (b) exhibits large-scale electronic localization as well.}
\label{fig0}
\end{center}
\end{figure}

\begin{figure}
\begin{center}
 \includegraphics[scale=0.53,trim={0 3cm 0 0},clip]{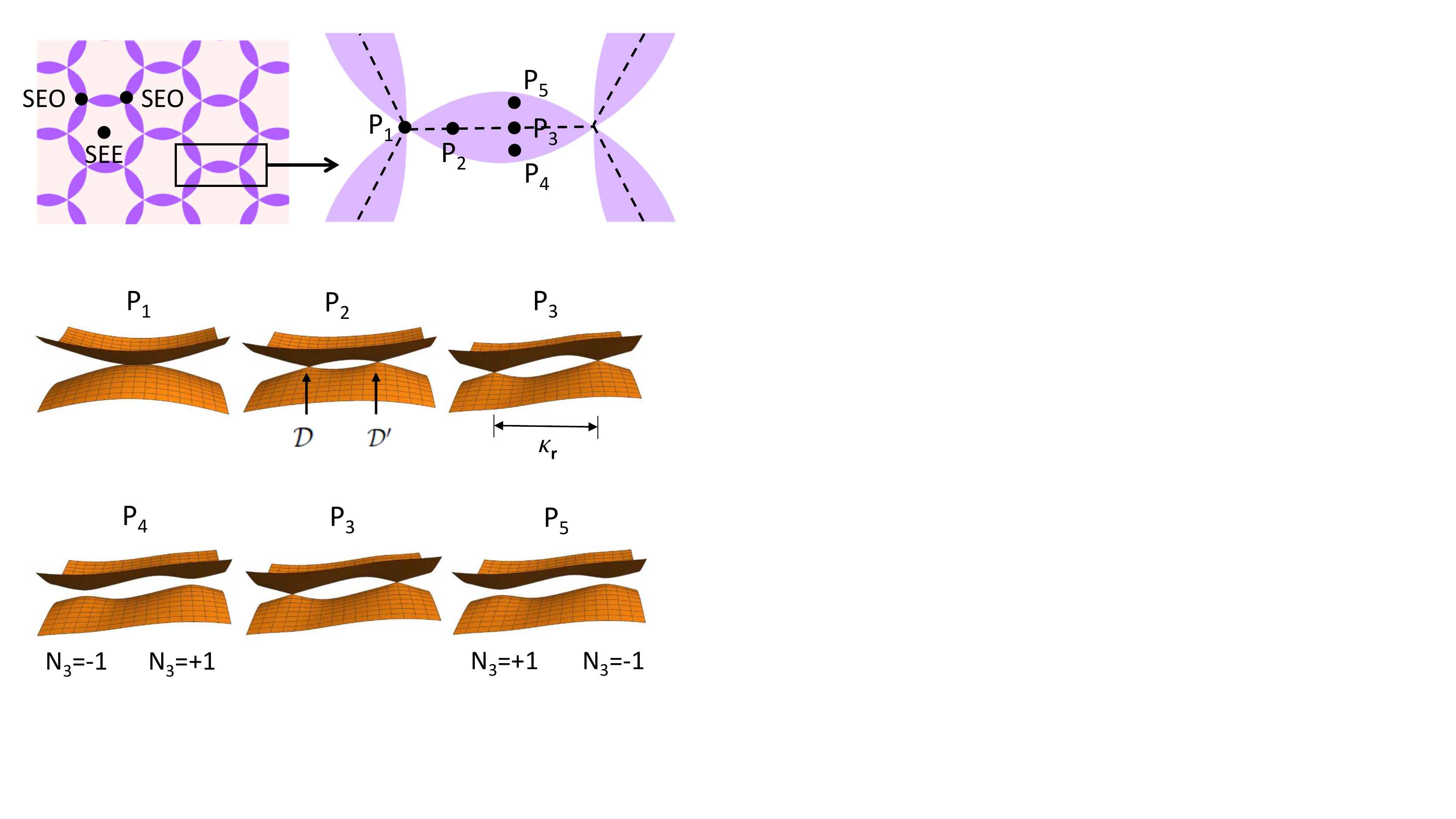}
\caption{As the angle of rotation approaches  the commensurate angle in Fig.~\ref{fig0}(b), the system separates into locally gapped (violet) and ungapped (off-white) regions. Semiclassical, local band structures at different points of the ungapped region are shown in the lower panels. Along the line joining nearest SEO points, two Dirac cones, $\mathcal{D}$ and $\mathcal{D}'$, appear. Their separation $\kappa_{\mb{r}}$ is zero at the SEO points and maximum midway between the SEO points. Along the perpendicular direction, the individual Dirac cones become gapped, even though in the violet regions there is no overall gap. The gap changes sign as one crosses the line between SEO points, with opposite sign change at the two Dirac points.}
\label{fig1}
\end{center}
\end{figure}

In incommensurate structures all such nontrivial interlayer effects so far have been looked for only at small rotation angles. In this Letter we show that,  intriguingly, they occur also at large ones---a regime that was previously thought to be trivial and featureless. This may sound counterintuitive since, at large angles, the moir{\'e} period is expected to be short, which suppresses the effect of  interlayer coupling. However, elementary geometric considerations show that large-scale moir{\'e} patterns appear not only at small angles, but also at large angles when the system is close to commensuration. In fact, small angles are merely a special case of such near commensurate structures: they appear near the `trivial' commensuration of zero angle.  
While near zero angle  moir{\'e} patterns interpolate between locally AA- and AB-stacked regions and are evident in the crystal structure,  near other commensurations, nearly SEE and SEO regions repeat  periodically, which is virtually imperceptible to the unguided eye, as demonstrated in Fig.~\ref{fig0}. However, being gapped and ungapped, respectively, SEE and SEO structures have qualitatively dissimilar electronic properties \cite{meleprb,melejphys}. Near generic commensuration angles, one thus expects a mosaic of locally gapped and ungapped regions, as shown in Fig.~\ref{fig1}. Below we demonstrate that, sufficiently close to commensuration, this visually `hidden' physics indeed has profound consequences: sign changes of the expected semiclassical gaps induce a set of topologically protected counterpropagating chiral modes percolating throughout the system that, surprisingly, support a  flat band arising from geometric frustration. This implies strong correlation physics as observed in Refs.\ \cite{mott,supercon}, but with exact flat bands and a Kagome localization pattern.

Consider a graphene bilayer with layers 1 and 2 mutually rotated by an angle $\theta$, not necessarily small, but close to some angle $\theta_c$ where the system is exactly commensurate, i.e., $|\delta\theta|=|\theta-\theta_c|\ll|\theta|$. The Dirac points of the unrotated and rotated layers are located at the corners of their respective Brillouin zones, $\mathbf{K}$ and $\mathbf{K}_{\theta}$. 
Since a hexagonal lattice   also has  a hexagonal Brillouin zone,  commensuration in real space implies commensuration in the extended zone scheme of reciprocal space. Therefore, at $\theta_c$ there necessarily exist two reciprocal lattice vectors, $\mb{G}$ and $\mb{G}'_{\theta_c}$such that $\mb{K} +\mb{G}=\mb{K}_{ \theta_c}+\mb{G}'_{\theta_c}$. Define $\mb{K}_{ \theta}+\mb{G}'_{\theta}-\mb{K} -\mb{G}=\delta\mb{K} $, where $\mb{G}'_{\theta}$ is $\mb{G}'_{\theta_c}$ rotated by $\delta\theta$.
Clearly $\delta K\ll K$ since $|\delta\theta|\ll1$. In Ref.~\cite{Palarxiv} a long-wavelength description   for such a system was derived in terms of the vector $\delta\mb{K}$.  The Hamiltonian of the system was shown to comprise two parts: an intralayer part described by Dirac Hamiltonians ($\hbar=1$):
\beq
 H_{1}=H_2=-iv_F \boldsymbol{\sigma}_{ }\cdot\mathbf{\nabla}, 
\label{dirac} 
\eeq
where $v_F$ is the Fermi velocity and $\boldsymbol{\sigma} = ( \sigma_x,\sigma_y)$ is a vector of Pauli matrices, and an  interlayer part depending on a Fourier component $\tilde{t}(\mb{q})$ of the coupling $t(\mathbf{\delta r})$ between atoms in different layers at lateral distance $\mathbf{\delta r}$: 
\begin{equation}
H_{\perp}(\mb{r})=\frac{\mathcal{V}}{3}\sum_{n=0}^2e^{i \delta\mathbf{K}_n\cdot\mathbf{r}}
\begin{pmatrix}
e^{-i \vartheta/2}&e^{-i 2\pi n/3}\\
e^{i 2\pi n/3}&e^{i \vartheta/2}
\end{pmatrix},
\label{hamreal}
\end{equation}
where $\mathcal{V}=\tilde{t}(\mb{K}+\mb{G})$, $\delta \mb{K}_n$ is $\delta\mb{K}$ rotated by $2n\pi/3$, and $\vartheta = \theta - 4\pi (l_1+l_2)/3$, where $l_{1,2}$ are coefficients that express $\mb{G}$ in terms of the reciprocal lattice vectors $\mb{b}_{1,2}$: $\mb{G}=l_1\mb{b}_1+l_2\mb{b}_2$.

When the superlattice is large, a semiclassical picture is justified, where one can consider the band structure locally at a given point in the superlattice, as shown in Fig.~\ref{fig1}. At $\mb{r}=0$ and its superlattice translates, the bandstructure resembles that of an $AA$-stacked bilayer, but with a gap arising from the difference between the diagonal terms in Eq.~(\ref{hamreal}). These are the SEE regions. On the other hand, at $\mb{r}= (4\pi/3\sqrt{3} \delta K^2) {\rm R}(\pm \pi/6) \mathbf{\delta K}$, where $R(\varphi)$ denotes a rotation by angle $\varphi$, and all superlattice translates, the local bandstructure is that of an AB (BA)-stacked bilayer without any gap. These are the SEO regions. In addition to the SEO points the regions connecting them are found to be ungapped as well, surrounding the gapped regions (see supplementary materials). Note that this is a property of systems near nonzero commensurate angles. Near the trivial commensuration $\theta_c=0$ one has $\vartheta\rightarrow 0$, and one recovers the Hamiltonian first derived in Ref.~\cite{lopes}, which does not produce any semiclassical local  gaps.

\begin{figure*}
\begin{center}
 \includegraphics[scale=0.5,trim={0 9.5cm 0 0},clip]{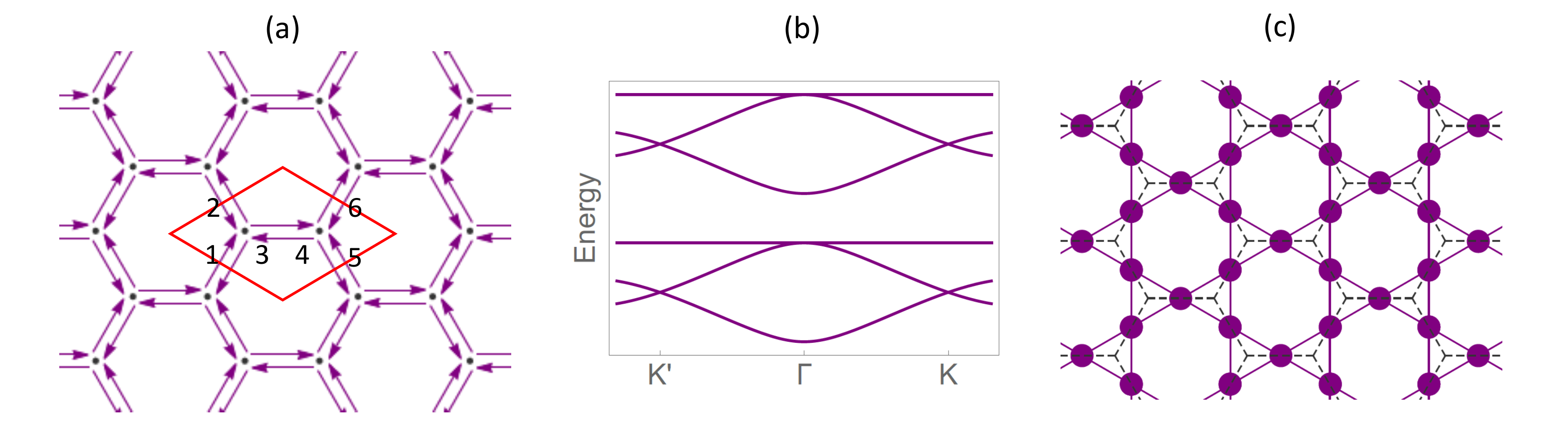}
\caption{(a) The network model corresponding to Fig.~\ref{fig1} with the unit cell shown in red. Incoming chiral modes scatter at the SEO points---called nodes---and leave as outgoing modes along pathways connecting two nodes---called links. (b) Energy spectrum of the network model: dispersive bands and flat bands appear, as in the spectrum of a Kagome lattice. (c) In the limit where   incoming modes backscatter strongly at the nodes, the states will be localized midway on each link. The emergent lattice of midpoints is a Kagome lattice. The system described by Fig.~\ref{fig1} can thus be effectively described as a tight-binding model on an emergent Kagome lattice, explaining the emergent band structure in (b).}
\label{fig2}
\end{center}
\end{figure*}

Let us consider the ungapped regions illustrated in Fig.~\ref{fig1} in more detail. The semiclassical, low-energy band structure consists of a pair of Dirac cones, $\mathcal{D}$ and $\mathcal{D}'$. They are not to be confused with the two inequivalent valleys of the single layer, $K$ and $K'$---both the cones  $\mathcal{D}$ and $\mathcal{D}'$ appear at the same valley and result from the interlayer coupling. Their separation, $\mb{\kappa}_{\mb{r}}$, is maximum midway between the two inequivalent SEO points, going to zero at the SEO points, where they merge to give rise to a parabolic touching point between the valence and conduction bands. Each Dirac cone is ungapped  on the line joining the SEO points. In the vicinity of this line a gap opens at each Dirac point, although  the band structure overall remains ungapped in the lobes seen in Fig.~\ref{fig1} due to indirect overlap \cite{footnote}. 
As one crosses this line, the gaps at the two Dirac points change sign, giving rise to two 1D chiral modes of opposite chirality along the line. Indeed, this is guaranteed for topological reasons \cite{vol}. Starting with Eqs.~(\ref{dirac}) and (\ref{hamreal}) and integrating out the high energy modes, the $2\times 2$ low-energy effective Hamiltonian truncated to linear order in $\mb{k}$ can be written as $h^\eta=\mb{g}^{\eta}(\mb{k})\cdot\mb{\sigma}$, where $\eta=\pm 1$ distinguishes the two Dirac cones  $\mathcal{D}$ and $\mathcal{D}'$. Calculating the topological charge $N_3^{\eta}=\int d\mb{k}\, \mb{g^\eta}\cdot(\partial_{k_x}\mb{g^\eta}\times\partial_{k_y}\mb{g^\eta})/4\pi|\mb{g^\eta}|^3$, we find that the difference of charges across the lines connecting SEO points is  $\eta$.
Although the topological charge sums up to zero for the two points together, in the absence of scattering between the two Dirac cones, one can consider each point separately. Thus, in the strong coupling limit, $\mathcal{V}/v_F\delta K\gg 1$, the corresponding index theorem implies localized modes---a pair of topologically protected 1D chiral modes---percolating through the system along the lines joining the SEO points. Similar arguments were used to predict---and observe---1D chiral modes at domain walls between AB- and BA-bilayer graphene gapped by an external electric field \cite{domain-theo,jose,domain-exp,huang,rickhaus}. In contrast, here these modes arise \emph{intrinsically} without requiring any external field.

To access the low-energy band structure, we construct a network model for these pairs of 1D chiral modes. In the long-wavelength limit and sufficiently far from the SEO points,  the two Dirac cones  $\mathcal{D}$ and $\mathcal{D}'$ are well-separated in momentum space, and there is no scattering between them. Each mode evolves freely until it reaches an SEO point, where the Dirac cones merge and scattering between them occurs. The SEO  points, thus, are the nodes of this network and  the lines connecting them links. As shown in Fig.~\ref{fig2}(a), each node has 3 incoming and 3 outgoing (chiral) modes. Because the lattice of nodes is hexagonal, pairs of adjoining nodes are inequivalent. Each is indexed by a tuple of integers $m, n$. We collect all the amplitudes for electrons to occupy the incoming and outgoing modes at a pair of lattice sites $m, n$ into vectors $|a_{mn}\rangle\equiv\{a_{1mn},...,a_{6mn}\}$ and $|b_{mn}\rangle\equiv\{b_{1mn},...,b_{6mn}\}$, respectively. The two can be related by a $6\times 6$ unitary matrix $\mathcal{U}$ describing the scattering between them at the nodes: $|b_{mn}\rangle=\mathcal{U}|a_{mn}\rangle.$
Each mode acquires a phase $e^{i\varepsilon}$ as it travels from one node to another; therefore, the incoming states at one node are related to the outgoing states in an adjoining node as $|a_{mn}\rangle_j=e^{-i\varepsilon}|b_{m+s(j),n+t(j)}\rangle_j$,
where $s(j), t(j)\in \{-1,0,1\}$ depend on geometry. Now applying Bloch's theorem, this can be recast into $|a(\mb{k})\rangle=e^{-i\varepsilon}\mathcal{M}(\mb{k})|b(\mb{k})\rangle$. Finally, with $|b(\mb{k})\rangle=\mathcal{U}|a(\mb{k})\rangle$ we obtain 
\beq
\mathcal{S}(\mb{k}) |b(\mb{k})\rangle=e^{i\varepsilon}|b(\mb{k})\rangle,
\label{eigen}
\eeq
where $\mathcal{S}(\mb{k})=\mathcal{U}\mathcal{M}(\mb{k})$. Eq.~(\ref{eigen}) is an eigenequation, and the phase $\varepsilon$ of its eigenvalues yields the energy of modes  $E = \varepsilon  {v}/L$, where $L$ is the length of the link and ${v}$ the mode velocity.

In order to solve Eq.~(\ref{eigen}) for $\varepsilon$ we need an expression for $\mathcal{U}$. We first note that $\mathcal{U}$ is block diagonal with entries $\mathcal{U}_+$ and $\mathcal{U}_-$, which are $3\times 3$ unitary matrices representing scattering at the two inequivalent nodes, respectively. 
The Hamiltonian possesses several symmetries: $C_3$ symmetry around each node, mirror reflection symmetry on the line joining the nodes, and point reflection on the midpoint of the line joining the nodes. Using these symmetries, we have (see supplementary materials)
\beq
\mathcal{U}_+=\mathcal{U}_-=e^{i\varphi}
\begin{pmatrix}
\alpha&\beta e^{i \lambda}&\beta e^{i \lambda}\\
\beta e^{i \lambda}&\alpha&\beta e^{i \lambda}\\
\beta e^{i \lambda}&\beta e^{i \lambda}&\alpha
\end{pmatrix},
\label{layerham}
\eeq
with $\alpha=1/\sqrt{1+8\mathrm{cos}^2\lambda}$ and $\beta=-2\mathrm{cos}\lambda/\sqrt{1+8\mathrm{cos}^2\lambda}$. The phase $e^{i\varphi}$ can be gauged out, leading to a one parameter model. Using Eq.~(\ref{layerham}) in Eq.~(\ref{eigen}) yields the exact energy spectrum,
\begin{eqnarray}
\varepsilon_{1}(\mathbf{k})&=&\begin{cases}
\lambda\pm\frac{1}{2}\mathrm{cos}^{-1}\left[\frac{(2c_{\mathbf{k}}-1)(1+\mathrm{cos}2\lambda)-1}{(5+4\mathrm{cos}2\lambda)}\right], \\
\mathrm{tan}^{-1}\left[\frac{\mathrm{sin}2\lambda}{2+\mathrm{cos}2\lambda}\right],
\end{cases}\label{energy1}\\
\varepsilon_{2}(\mathbf{k})&=&\varepsilon_{1}(\mathbf{k})+\pi,
\label{energy2}
\end{eqnarray}
where $c_{\mb{k}}=2\mathrm{cos}\left(\frac{\sqrt{3}k_yL}{2}\right)\mathrm{cos}\left(\frac{3k_xL}{2}\right)+\mathrm{cos}\left(\sqrt{3}k_yL\right)$.
The spectrum consists of a pair of triplets with each triplet comprising two dispersive bands and a flat band, as shown in Fig.~\ref{fig2}(b). The dispersive bands intersect each other linearly at Dirac points and the flat band touches one of the dispersive bands. Also, from Eq.~(\ref{eigen}) it follows that if $\varepsilon$ is a solution, so is $\varepsilon+2N\pi$, where $N$ is any integer, i.e., the sextet pattern repeats periodically in energy. The scattering parameter $\lambda$   only affects the bandwidth, but not the qualitative shape of the bands. 

The above band diagram is strongly reminiscent of the tight-binding bands on a Kagome lattice. The question arises whether this similarity is accidental or whether  there is a deeper link. To elucidate, consider a limiting situation where each incoming mode back scatters very strongly at the nodes, i.e., $\lambda\rightarrow\pi/2$, localizing electrons on the links. The system is then described effectively by a tight-binding model on a lattice with sites at the centers of network links, as shown in Fig.~\ref{fig2}(c). This lattice is indeed a Kagome lattice.   Guided by this clue it follows readily that localized modes and the corresponding  flat bands arise in our network model---independent of the scattering parameter $\lambda$---from geometric frustration just as on the Kagome lattice. This frustration built into the geometry of localization makes the flat bands discussed here robust to moderate variations in experimental conductions, such as the rotation angle. In contrast, the localization probed in Refs.\ \cite{mott,supercon} has a triangular pattern and is complete only when precisely at some  'magic' angles \cite{bistritzer}.

\begin{figure}
\begin{center}
 \includegraphics[scale=0.52,trim={0 4cm 0 0},clip]{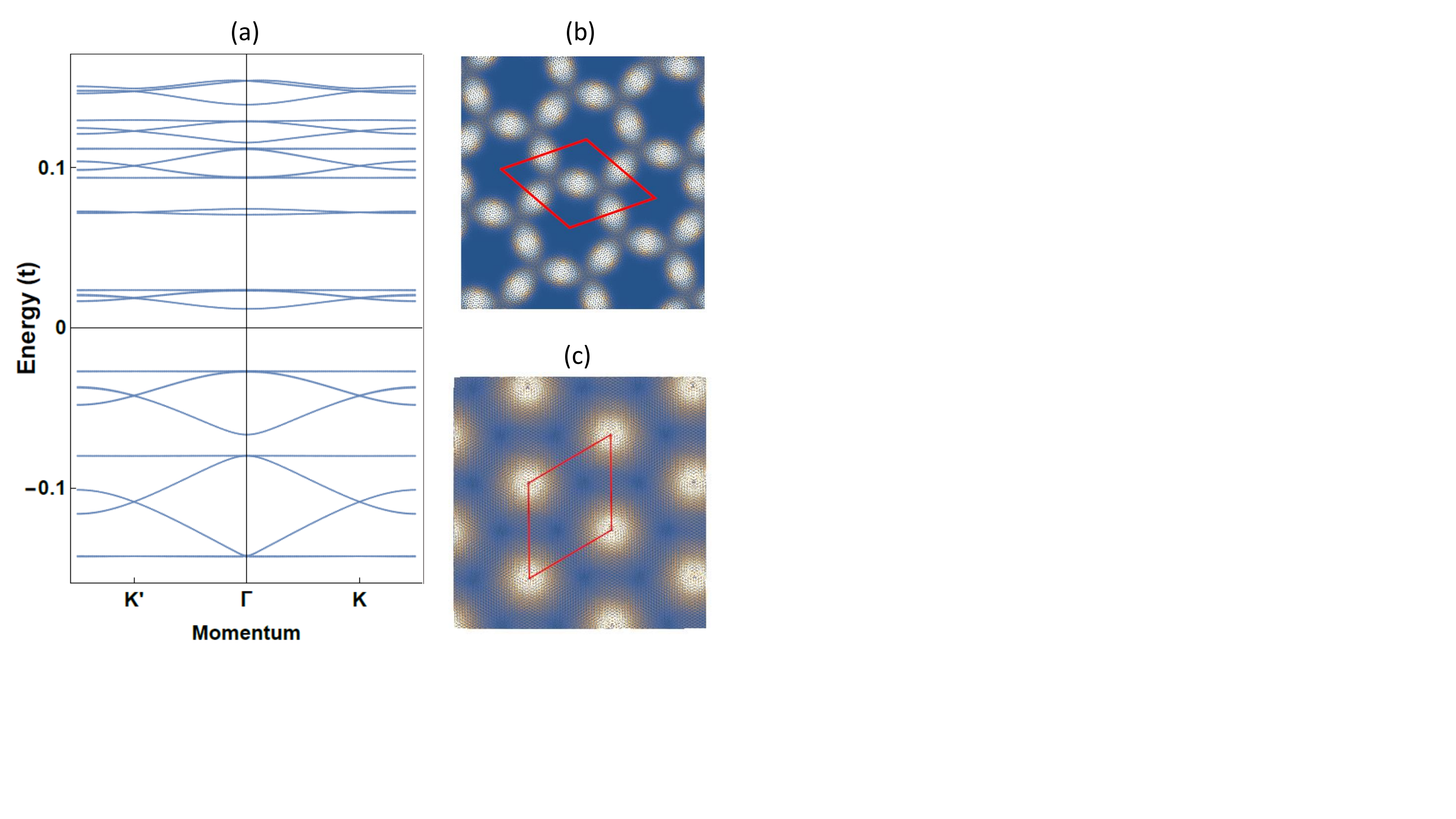}
\caption{(a) Numerically calculated low energy bandstructure  and (b) localization of electrons in the lowest energy band  of a system at $\theta = 38.546^{\circ}$. The interlayer coupling is described in the text. Both the band structure and the localization pattern are as predicted by our low-energy network model [cf. Figs.~\ref{fig2}(b) and (c)]. (c) Localization of electrons at $\theta = 1.47^{\circ}$. Electrons in the large angle and small angle regimes localize in  Kagome (b) and triangular (c) patterns, respectively, giving access to different correlation physics.}
\label{fig3}
\end{center}
\end{figure}

We now test the above predictions by directly diagonalizing tight-binding (TB) models of moir{\'e} graphene bilayers. The emergent frustration is expected in the strong coupling regime $\mathcal{V}/v_F\delta K\gg 1$. We thus choose the commensuration with   $\theta_c=38.21^\circ$, which has the largest $\mathcal{V}$ of all    $\theta_c\neq 0$. In that case it is estimated $\mathcal{V}\approx 10\mathrm{meV}$ \cite{shallcross} and the localization length of the 1D modes $v_F/\mathcal{V}$ with $v_F=10^6\mathrm{m/s}$ thus  becomes $\sim10^3\AA$ or $\sim 700$ lattice constants. Therefore  a lattice with $\approx10^6$ atoms is needed. Performing TB calculations on such a large system is a formidable task. In order to make numerical calculations feasible, we reduce the required lattice size by artificially enhancing $\mathcal{V}=\tilde{t}(\mb{K}+\mb{G})$, choosing an interlayer coupling $t(\mathbf{\delta r})=t_{\rm art}(\mathbf{\delta r})$ with dominant Fourier component at the momentum $\mb{K}+\mb{G}$.

The results below are for a graphene bilayer with interlayer rotation angle $\theta=38.546^\circ$. 
We choose the artificial coupling to be a Bessel function: $t_{\rm art}(\mb{\delta r})=t_{1} J_0(G\delta r)\theta_{\rm H}(\delta r-l_0)$, where $t_{1} =0.2t$ in terms of the intra-layer  hopping energy $t$, $l_{0}=6a$ in terms of the nearest carbon-carbon distance $a$, $G = 4\pi\sqrt{7}/3a$, and $\theta_{\rm H}$ is the Heaviside step function.  We have checked  that this indeed produces an enhanced $\mathcal{V}$.
 Fig.~\ref{fig3}(a) shows the band structure of this TB model. Appearance of flat bands along with dispersive ones is evident. As predicted by our network model, at low energies, the flat band touches one of the dispersive bands while the two dispersive bands intersect themselves at Dirac points. Moreover, the pattern is periodic along the energy axis. Plotting the electron density in Fig.~\ref{fig3}(b), we find that the electrons are localized on the links connecting the SEO regions,  
forming a Kagome pattern. Thus, all the main features predicted by our analytic theory are supported by TB calculations. For comparison, we have also plotted in Fig.~\ref{fig3}(c) the localization pattern near the first magic angle. The pattern, well documented in previous works \cite{laissardiere}, is triangular.

In constructing our low-energy theory we assumed the incoming and outgoing modes to scatter only at the nodes [Fig.~\ref{fig2}(a)]. When not in the long-wavelength limit, such as in our TB calculations, this assumption does not hold and electrons may scatter as they travel along the links as well. 
By extending our network model correspondingly, we find that extra scattering in the middle of each link produces qualitatively new features: the particle-hole symmetry in the dispersive bands is lost and a pair of triplets may break up into a quartet and a doublet---see supplementary materials. On closer inspection of the TB band diagram in Fig.~\ref{fig3}(a), we find these new qualitative features present as well; therefore, we ascribe them to such extra scattering. Importantly, the flat bands persist nevertheless, and are immune to such modifications.

We emphasize, however, that in the long-wavelength limit of large superlattice size $L$, inaccessible to our numerics,  such extra scattering is absent and the results in Eqs.~(\ref{energy1}) and (\ref{energy2}) become exact. This can be seen as follows: 
Referring to Fig.~\ref{fig1}, to avoid scattering between the two modes on a link, the momentum spread in the semiclassical wavefunctions, $\Delta p$, must be smaller than the distance $\kappa_F$ between the respective Fermi points of the Dirac cones $\mathcal{D}$ and $\mathcal{D}^{\prime}$,    i.e., $\Delta p \ll \kappa_F $. At zero chemical potential and near the SEO regions we find $\kappa_F \sim (\mathcal{V}/v) \sqrt{2x/L}$, where $x$ is the distance from the closest SEO point along the link. On the other hand, the wavepacket spread scales as $\Delta p \sim 1/\Delta y$, where $\Delta y$ is set by the size of the semiclassically allowed region [the lobes between SEO points in Fig.~\ref{fig1}]: $\Delta y \sim 2 x |\vartheta-\pi|/3$ (valid at $x \ll L$). Combining the two, the condition $\Delta p \ll \kappa_F $ yields
$x/L \gg (v/|\vartheta-\pi|\mathcal{V}L)^{2/3}$. It means that  in the limit of  large $L$, the region where scattering occurs shrinks to zero relative to the length of the links. Scattering then occurs only at the nodes of our network model. Eqs.~(\ref{energy1}) and (\ref{energy2}) are, thus, the exact solutions to the low-energy band structure of   graphene bilayers sufficiently close to nonzero commensuration angles.

In conclusion, moir{\'e} graphene bilayers exhibit intriguing phenomena in a parameter regime previously considered perturbative and uninteresting: large, incommensurate rotation angles. We have demonstrated that near commensuration electrons are channeled through the system along a geometrically frustrated network of topologically protected modes. The consequent electron localization implies strong correlations  with the exciting prospect of exploring spin-liquid physics and other exotic states of matter on a Kagome lattice in this highly tunable system \cite{exo1,exo2,exo3}. Feasibility of observing such correlated states in twisted graphene bilayers has just been demonstrated  \cite{mott,supercon}. Based on the estimate of  $\mathcal{V}$ in Ref.\ \cite{shallcross} we expect that observation of the predicted effects requires  twist angles in an interval of width on the order of  $0.05^\circ$  around the commensurate angle  $\theta_c=38.21^\circ$, an accuracy approached by the experiments of Refs.\ \cite{kim-nano,mott,supercon}.  The emergent geometric frustration allows to explore  exactly flat bands over that entire  angular range. In addition, the non-interacting low-energy theory near commensurations has an exact analytical solution. This will facilitate  theoretical investigation  of the consequent strongly correlated physics \cite{topostrong,kim,gonzalez}. We remark that the physics described here is not restricted only to graphene but readily generalizes to other bilayer sandwiches of Dirac materials \cite{palprb,othervdw1,othervdw2}. 

\begin{acknowledgements}
We acknowledge support by NSF under DMR-1055799.
\end{acknowledgements}

\begin{widetext}

\section{Supplementary Materials}

\section{Gapped and ungapped regions in the superlattice}

The Hamiltonian of a graphene bilayer with a mutual angle of rotation $\theta$ near some commensuration angle $\theta_c$ is given by
\beq
H=
\begin{pmatrix}
H_1&H_{\perp} \\
H_{\perp}^{\dagger}&H_2
\end{pmatrix},
\label{layerham_supp}
\eeq
 where $H_{i}$ is the intralayer Hamiltonian of layer $i$  and $H_{\perp}$ couples the layers. 
As stated in the main text, in the continuum approximation, we have
\beq
 H_{1}=H_2=H_0=-iv_F \boldsymbol{\sigma}_{ }\cdot\mathbf{\nabla}, 
\label{dirac_supp} 
\eeq
where $v_F$ is the Fermi velocity, $\boldsymbol{\sigma} = ( \sigma_x,\sigma_y)$ is a vector of Pauli matrices acting on the sublattice space, and we set $\hbar=1$; and, the interlayer part depending on the Fourier component $\tilde{t}(\mb{q})$ of the coupling $t(\mathbf{\delta r})$ between atoms in different layers at lateral distance $\mathbf{\delta r}$, 
\begin{equation}
H_{\perp}(\mb{r})=\frac{\mathcal{V}}{3}\sum_{n=0}^2e^{i \delta\mathbf{K}_n\cdot\mathbf{r}}
\begin{pmatrix}
e^{-i \vartheta/2}&e^{-i 2\pi n/3}\\
e^{i 2\pi n/3}&e^{i \vartheta/2}
\end{pmatrix},
\label{hamreal_supp}
\end{equation}
where $\mathcal{V}=\tilde{t}(\mb{K}+\mb{G})$, $\delta \mb{K}_n$ is $\delta\mb{K}$ rotated by $2n\pi/3$, and $\vartheta = \theta - 4\pi l/3$, where $l=l_1+l_2$, with $l_{1,2}$ as the coefficients expressing $\mb{G}$ in terms of the reciprocal lattice vectors $\mb{b}_{1,2}$: $\mb{G}=l_1\mb{b}_1+l_2\mb{b}_2$.

Assuming that the supercell is much larger than the single layer lattice spacing, the Hamiltonian changes sufficiently slowly in real space so that, at any given point in real space, one can locally go to the momentum space and construct a band structure. Thus, $H_0$ in Eq.~(\ref{dirac_supp}) is replaced by $\mb{\sigma}\cdot\mb{k}$ and $\mb{r}$ in Eq.~(\ref{hamreal_supp}) is treated as a parameter. The Hamiltonian leads to a band diagram that is gapped in certain regions, but ungapped in others. The local Hamiltonians Eq.~(\ref{layerham_supp}) have a chiral symmetry coupled with inversion symmetry $\Sigma = l_z P_{\bf k}$, where we defined $\mb{l}=(l_x,l_y,l_z)$ as a vector of Pauli matrices acting on the layer space and $P_{\bf k}$ is point reflection at the point of zero momentum (here, in the semiclassical Hamiltonians, naturally the parameter $\mb{r}$ is not inverted, but only the momentum $\mb{k}$). Regions with no gap thus require a zero energy eigenstate. This means that the spectrum is gapped if and only if Det$[H]=0$. After some straightforward but tedious algebra one can show that this condition reduces to
\beq
b+\sqrt{2}c\le 0,
\eeq
where $c=\frac{\mathcal{V}^4}{9}\sum_{n,n'}\mathrm{cos}[(\delta\mb{K}_n-\delta\mb{K}_{n'})\cdot\mb{r}]$, $b=b_0-b_1$, with $b_0=-\frac{2c}{\mathcal{V}^2}\mathrm{cos}(\theta+4\pi l/3)$ and $b_1=\frac{4\mathcal{V}^2}{9}\{\sum_n(\mathrm{cos}[\tilde{\delta\mb{K}}_n\cdot\mb{r}])^2-\frac{1}{2}\sum_{n\ne n'}\mathrm{cos}[\tilde{\delta\mb{K}}_n\cdot\mb{r}]\mathrm{cos}[\tilde{\delta\mb{K}}_n'\cdot\mb{r}]\}^{1/2}$. Here $\tilde{\delta\mb{K}}_n=\delta\mb{K}_n-\delta\mb{K}_{n+1}$. In Fig.~[\ref{fig1_supp}] we show regions satisfying this condition and contrast them with those that do not for two values of rotation angle $\theta$: the blue regions are ungapped while the remaining ones are gapped. The thickness of the regions depend on the angle of rotation $\theta$, but the basic structure remains the same. 

\begin{figure}
\centering
\subfigure[]{\includegraphics[width=.28\textwidth]{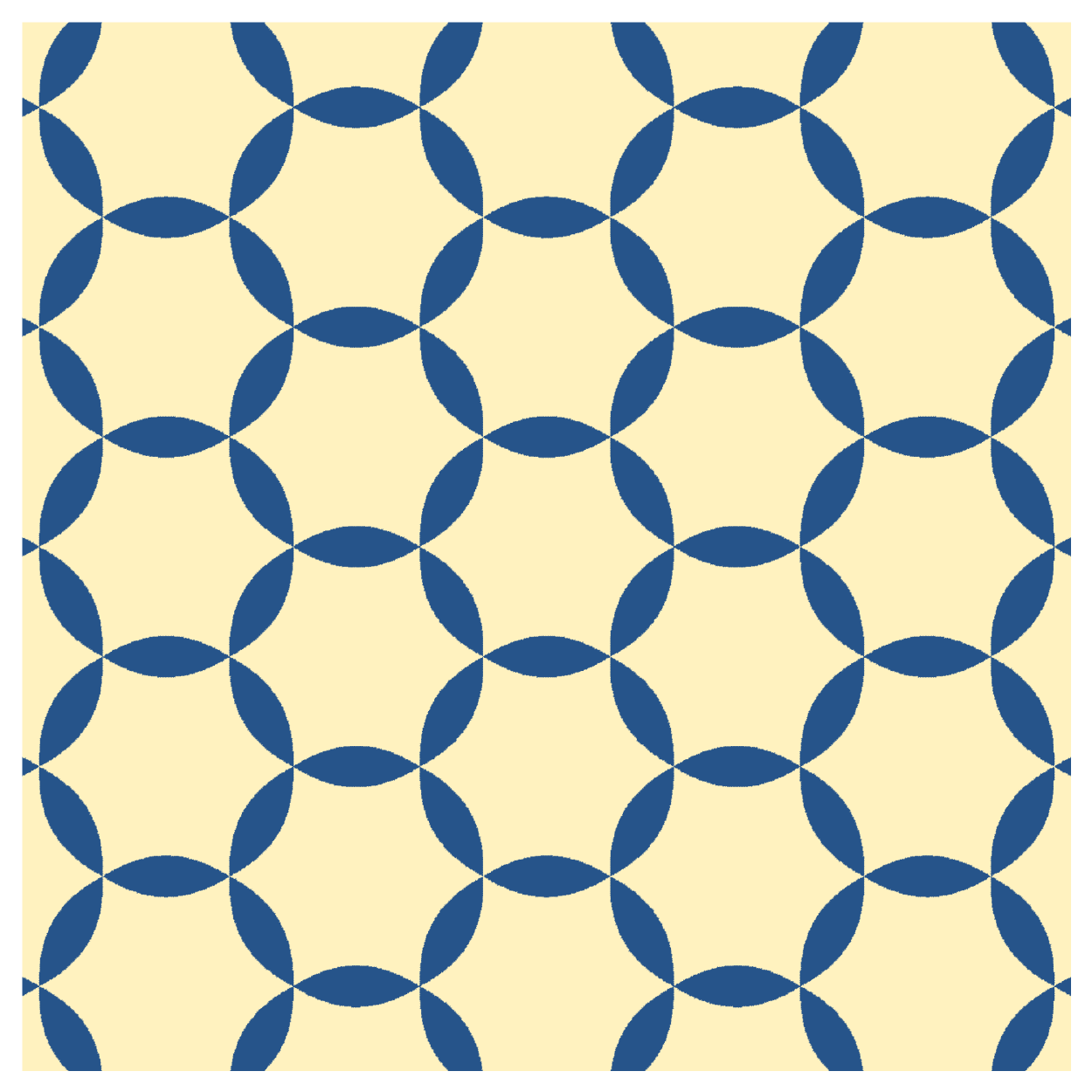}
\label{fig11_supp}}
\quad
\subfigure[]{\includegraphics[width=.28\textwidth]{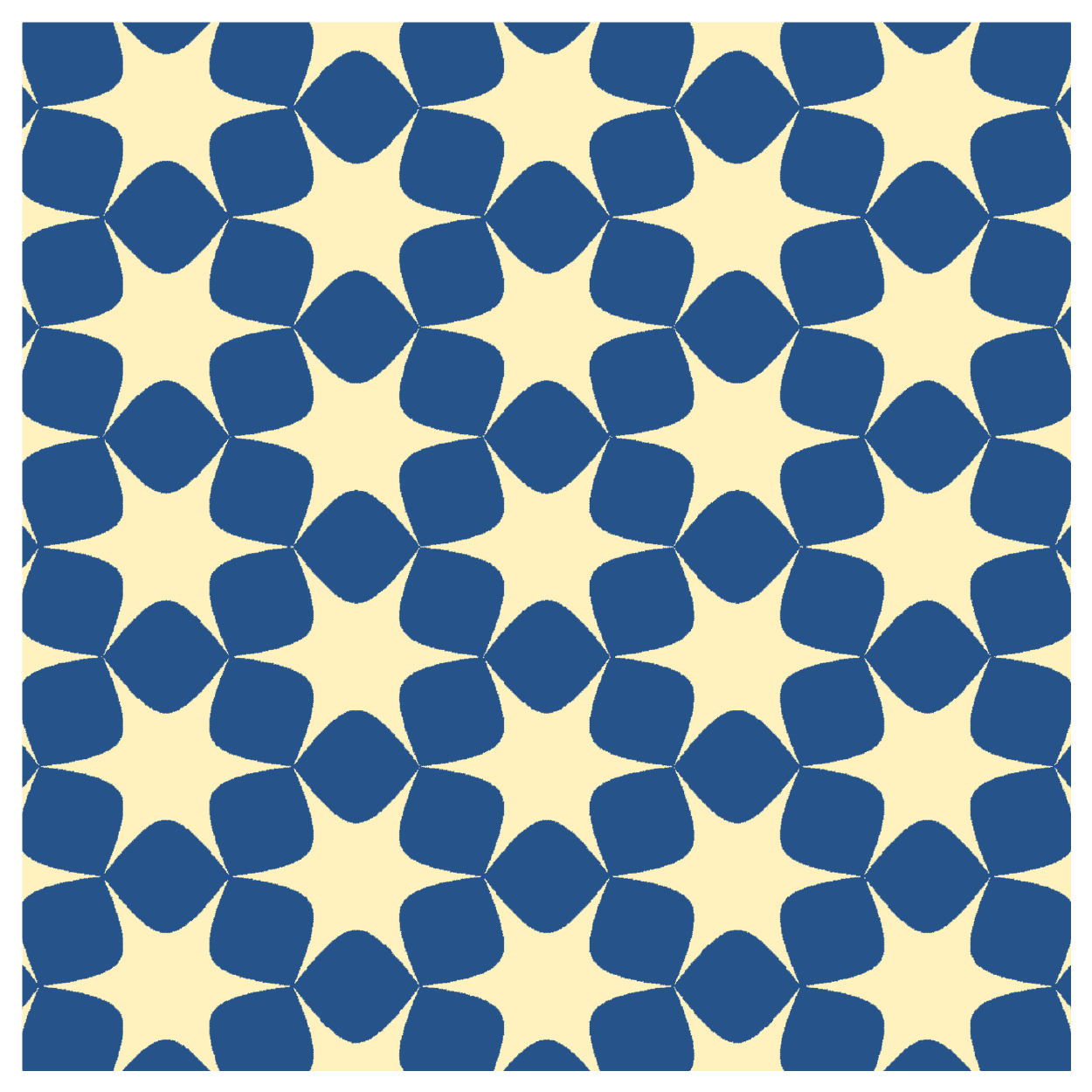}
\label{fig12_supp}}
\caption{Gapped (off-white) and ungapped (blue) regions in the superlattice for two values of $\theta$. While the thickness of the lobes (ungapped regions) changes with $\theta$, the basic structure remains unaltered.}
\label{fig1_supp}
\end{figure}

\section{Symmetries of the Hamiltonian and the resulting scattering matrix}

In order to understand the symmetries of the Hamiltonian in Eq.~(\ref{layerham_supp}), we rewrite it in a compact form:
\beq
H=-iv_F \boldsymbol{\sigma}\cdot\mathbf{\nabla}+\frac{\mathcal{V}}{3}\sum_{n=0}^2e^{il_z\delta\mb{K}_n\cdot\mb{r}/2}\left[e^{-il_z\sigma_z\vartheta/4}l_xe^{il_z\sigma_z\vartheta/4}+e^{-i\sigma_z n\pi/3 }l_x\sigma_xe^{i\sigma_zn\pi /3}\right]e^{-il_z\delta\mb{K}_n\cdot\mb{r}/2},
\label{hamcompact_supp}
\eeq
where we have used Eqs.~(\ref{dirac_supp}) and (\ref{hamreal_supp}) in Eq.~(\ref{layerham_supp}). The above Hamiltonian satisfies the following symmtries:
\begin{eqnarray}
\mathcal{C}_3&=&C_3e^{-i  l_z s \pi/3}e^{i \sigma_z\pi/3},\label{c3_supp}\\
\mathcal{M}&=&Ml_x\sigma_x,\label{m_supp}\\
\mathcal{P}&=&P\sigma_x\mathcal{K},\label{p_supp}
\end{eqnarray}
where $C_3$ denotes rotation by $2\pi/3$ around the two inequivalent nodes denoted by $s=\pm$ (see Fig. \ref{fig2_supp}), $M$ denotes mirror reflection on the line joining the two inequivalent nodes, $P$ denotes a point reflection on the midpoint between the two inequivalent nodes, and $\mathcal{K}$ denotes complex conjugation.

To prove the above symmetries, it is convenient to write $\mathbf{r}=\mb{r}_s+\mb{r}_{0s}$ in Eq.~(\ref{hamcompact_supp}), where $\mb{r}_{0s}$ defines the position of the two inequivalent nodes denoted by $s=\pm$ (see Fig.~\ref{fig2_supp}).  Because $e^{i \delta\mathbf{K}_n\cdot\mathbf{r}_{0s}}=e^{i s n 2 \pi/3}$ (up to a constant phase that can be gauged out), Eq.~(\ref{hamcompact_supp}) reduces to
\beq
H=-iv_F \boldsymbol{\sigma}\cdot\mathbf{\nabla}+\frac{\mathcal{V}}{3}\sum_{n=0}^2e^{il_z\delta\mb{K}_n\cdot\mb{r}_s/2}e^{i l_zs n  \pi/3}\left[e^{-il_z\sigma_z\vartheta/4}l_xe^{il_z\sigma_z\vartheta/4}+e^{-i\sigma_z n\pi/3 }l_x\sigma_xe^{i\sigma_zn\pi /3}\right]e^{-il_z s n  \pi/3}e^{-il_z\delta\mb{K}_n\cdot\mb{r}/2}.
\label{hamcompact2_supp}
\eeq
First, we consider the rotational symmetry $\mathcal{C}_3$. A rotation  $R(2\pi/3)$ by $2\pi/3$ around one of the nodes transforms $\mb{r}_s\rightarrow R(2\pi/3)\mb{r}_s$. This results in 
\begin{eqnarray}
e^{il_z\delta\mb{K}_n\cdot\mb{r}_s/2}&\rightarrow& e^{il_z \delta\mathbf{K}_n\cdot R(2\pi /3)\mathbf{r}_s/2}= e^{il_z \delta\mathbf{K}_{n-1}\cdot\mathbf{r}_s/2},\nonumber\\
\boldsymbol{\sigma}\cdot\mathbf{\nabla}&\rightarrow&e^{-i \sigma_z\pi/3}\boldsymbol{\sigma}\cdot\mathbf{\nabla}e^{i \sigma_z\pi/3}.\nonumber
\end{eqnarray}
Using this in Eq.~(\ref{hamcompact2_supp}), relabelling $n-1$ as $n'$, and multiplying the resulting $H$ with $e^{-i  l_z s \pi/3}e^{i \sigma_z\pi/3}$ from left and its inverse from the right, we recover the original Hamiltonian in Eq.~(\ref{hamcompact2_supp}).   $\mathcal{C}_3$ as given in Eq.~(\ref{c3_supp}) thus is a symmetry of $H$.

\begin{figure}
\begin{center}
  \includegraphics[scale=0.5,trim={0 5cm 0 1cm},clip]{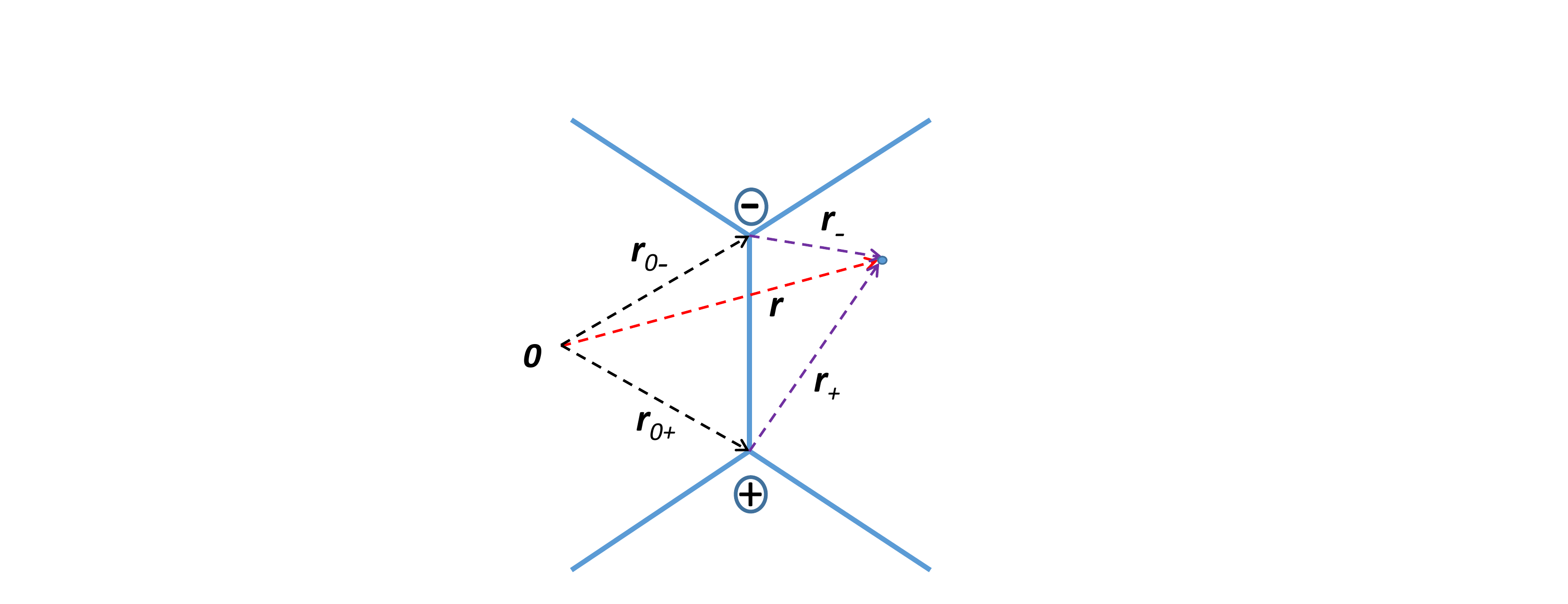}
  \caption{Diagram to understand the symmetries of the Hamiltonian in Eq.~\ref{layerham_supp}.}
\label{fig2_supp}
\end{center}
\end{figure}

Next, we investigate the mirror symmetry $\mathcal{M}$. A mirror reflection on the line joining the two inequivalent nodes transforms $\mb{r}_s=(x_s,y_s)\rightarrow (x_s,-y_s)$, where $x_s$ is taken to be along the line joining the nodes, and $y_s$ is perpendicular to it. This results in 
\begin{eqnarray}
e^{il_z\delta\mb{K}_n\cdot\mb{r}_s/2}&\rightarrow &e^{-il_z \delta\mathbf{K}_{-n}\cdot \mathbf{r}_s/2},\nonumber\\
\boldsymbol{\sigma}\cdot\mathbf{\nabla}&\rightarrow&-\boldsymbol{\sigma}^{\ast}\cdot\mathbf{\nabla}.\nonumber
\end{eqnarray}
Using this in Eq.~(\ref{hamcompact2_supp}), relabelling $-n$ as $n'$, and multiplying the resulting $H$ with $l_x\sigma_x$ from left and  right, we recover the original Hamiltonian in Eq.~(\ref{hamcompact2_supp}).  Therefore also  $\mathcal{M}$ as given in Eq.~(\ref{m_supp}) is a symmetry of $H$.

Finally, we come to the point inversion symmetry $\mathcal{P}$ that relates the two inequivalent nodes. A point reflection on the midpoint of the line joining the two nodes transforms $\mb{r}_s\rightarrow -\mb{r}_{-s}$. This results in
\begin{eqnarray}
e^{il_z\delta\mb{K}_n\cdot\mb{r}_s/2}&\rightarrow &e^{-il_z \delta\mathbf{K}_{n}\cdot \mathbf{r}_{-s}/2},\nonumber\\
\boldsymbol{\sigma}\cdot\mathbf{\nabla}&\rightarrow&-\boldsymbol{\sigma}\cdot\mathbf{\nabla}.\nonumber
\end{eqnarray}
Using this in Eq.~(\ref{hamcompact2_supp}), relabelling $-s$ as $s'$, taking the complex conjugate, and multiplying the resulting $H$ with  $\sigma_x$ from left and right, we recover the original Hamiltonian in Eq.~(\ref{hamcompact2_supp}).  This shows that also $\mathcal{P}$ as given in Eq.~(\ref{p_supp}) is a symmetry of $H$.

The $\mathcal{C}_3$ symmetry [Eq.~(\ref{c3_supp})] requires the scattering matrix for each node to be of the form
\beq
\mathcal{U}=e^{i\varphi}
\begin{pmatrix}
\alpha&\beta e^{i \lambda}&\gamma e^{i \omega}\\
\gamma e^{i \omega}&\alpha&\beta e^{i \lambda}\\
\beta e^{i \lambda}&\gamma e^{i \omega}&\alpha
\end{pmatrix}.
\label{scatmat_supp}
\eeq
Mirror symmetry $\mathcal{M}$  [Eq.~(\ref{m_supp})] reduces it to
\beq
\mathcal{U}=e^{i\varphi}
\begin{pmatrix}
\alpha&\beta e^{i \lambda}&\beta e^{i \lambda}\\
\beta e^{i \lambda}&\alpha&\beta e^{i \lambda}\\
\beta e^{i \lambda}&\beta e^{i \lambda}&\alpha
\end{pmatrix}.
\label{scatmat2_supp}
\eeq
The point reflection symmetry $\mathcal{P}$ [Eq.~(\ref{p_supp})] relates the scattering matrices at the two inequivalent nodes. Because of the complex conjugation, which reverses the direction of motion of electrons, this symmetry maps incoming modes into outgoing ones, relating the scattering matrix at one node to the inverse of the scattering matrix at the  other node as
\beq
\mathcal{U}_-=\mathcal{U}_+^{-1\ast}.
\eeq
Using the fact that $\mathcal{U}$ is unitary, together with $\mathcal{U}=\mathcal{U}^T$, as seen in Eq.~(\ref{scatmat2_supp}), we find
\beq
\mathcal{U}_-=\mathcal{U}_+=\mathcal{U}.
\label{twonodes_supp}
\eeq
Eq.~(\ref{scatmat2_supp}) along with Eq.~(\ref{twonodes_supp}) appear in the  main text.

\section{Effect of additional scattering in between two nodes}

In the limit of large system size, we expect scattering of the zero energy modes only at the network nodes.  In this case, the scattering matrix $\mathcal{U}$ is sufficient to describe the low energy electronic structure of the system.  However, small systems will exhibit complicated scattering along the network links.  Although we cannot hope for a full quantitative understanding of these processes in a our simple effective model, we can understand how they distort the electronic structure qualitatively.

We now examine how scattering on the midpoint of every link affects the band structure.  The  symmetries discussed above reduce also this process to one parameter.  We include this new scattering in the geometric matrix $\mathcal{M}$ via the parameter $\tau$:

\beq
\mathcal{M}(\mb{k}) \to \sin{(\tau)} \mathcal{I} + i\cos{(\tau)} \mathcal{M}(\mb{k}).
\eeq

\begin{figure}
\centering
\subfigure[]{\includegraphics[width=.4\columnwidth]{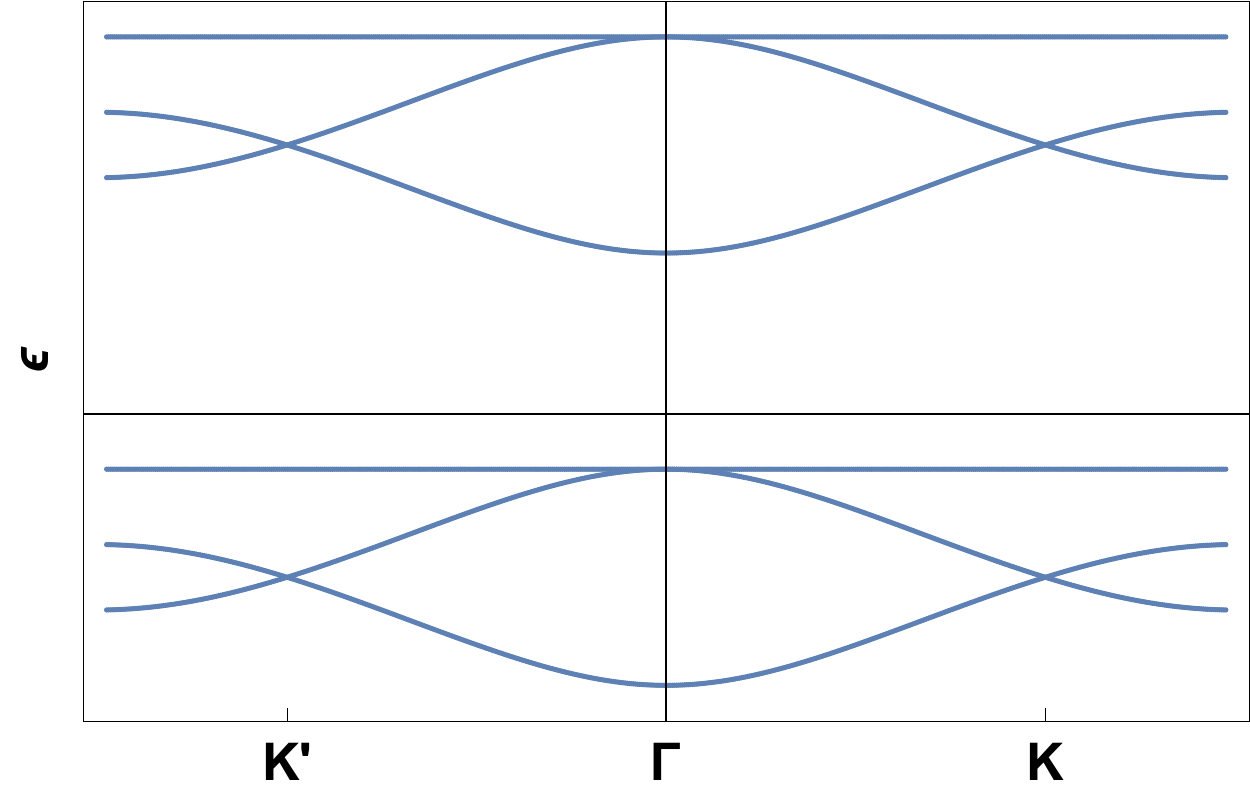}}
\quad
\subfigure[]{\includegraphics[width=.4\columnwidth]{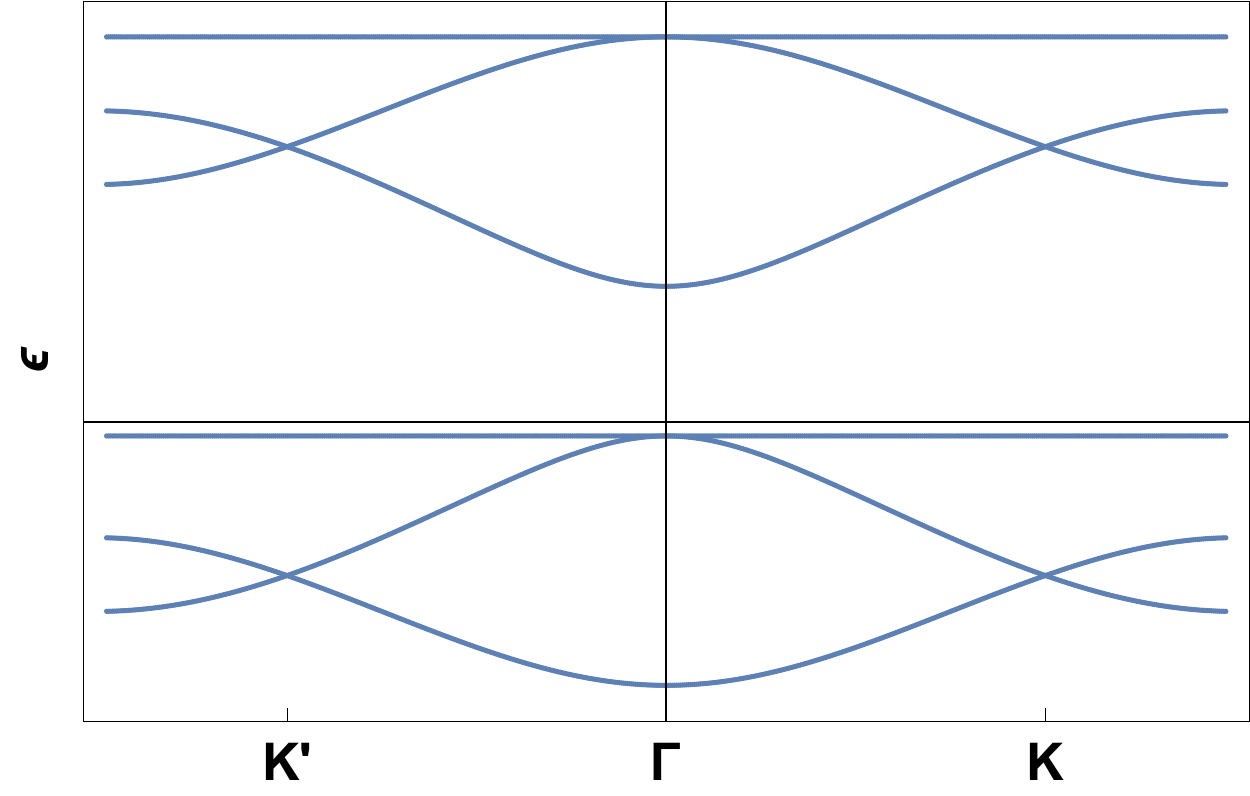}}
\quad
\subfigure[]{\includegraphics[width=.4\columnwidth]{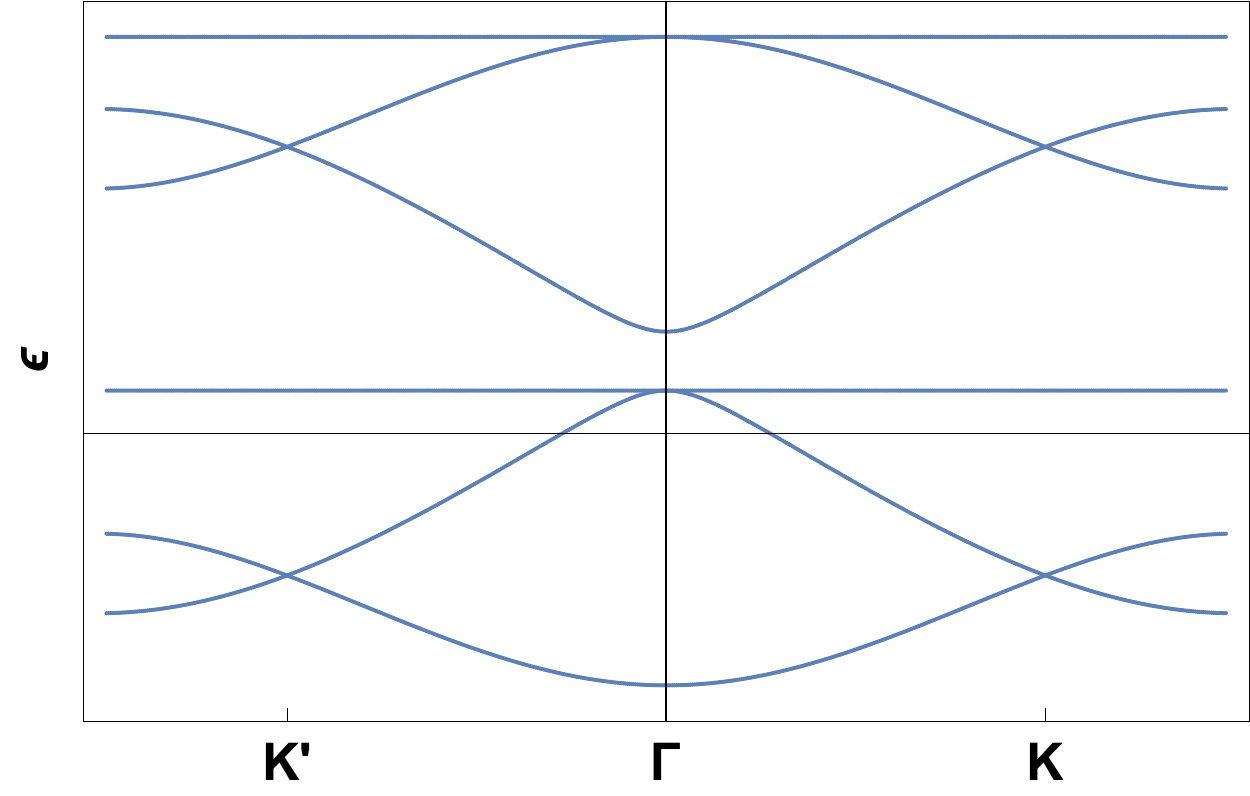}}
\quad
\subfigure[]{\includegraphics[width=.4\columnwidth]{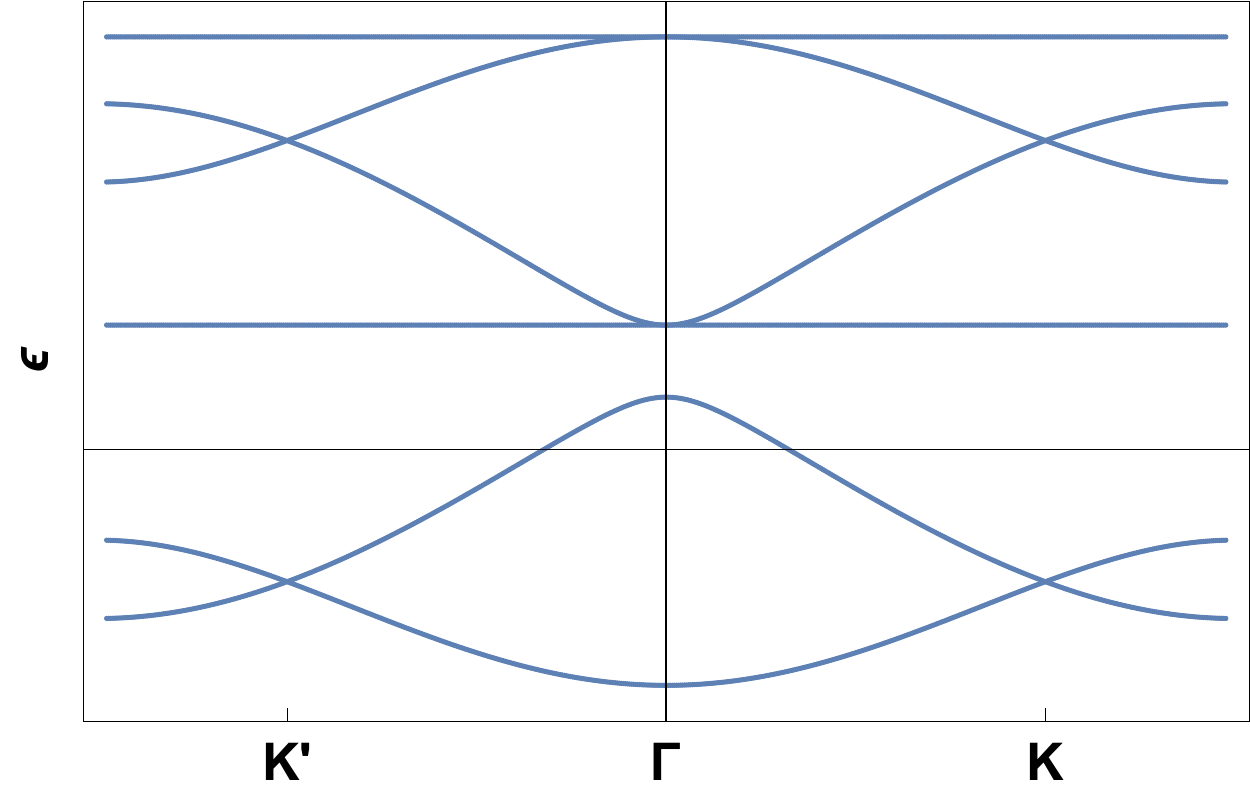}}
\quad
\subfigure[]{\includegraphics[width=.4\columnwidth]{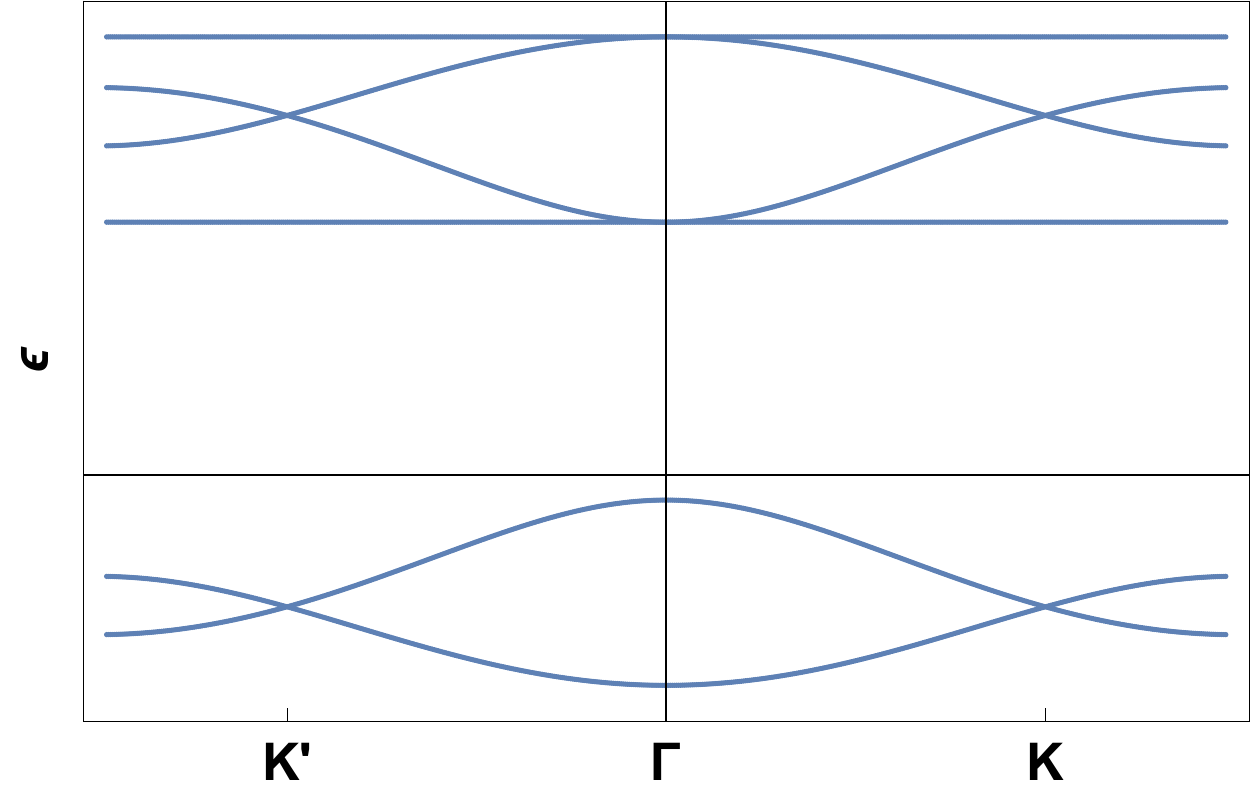}}
\quad
\subfigure[]{\includegraphics[width=.4\columnwidth]{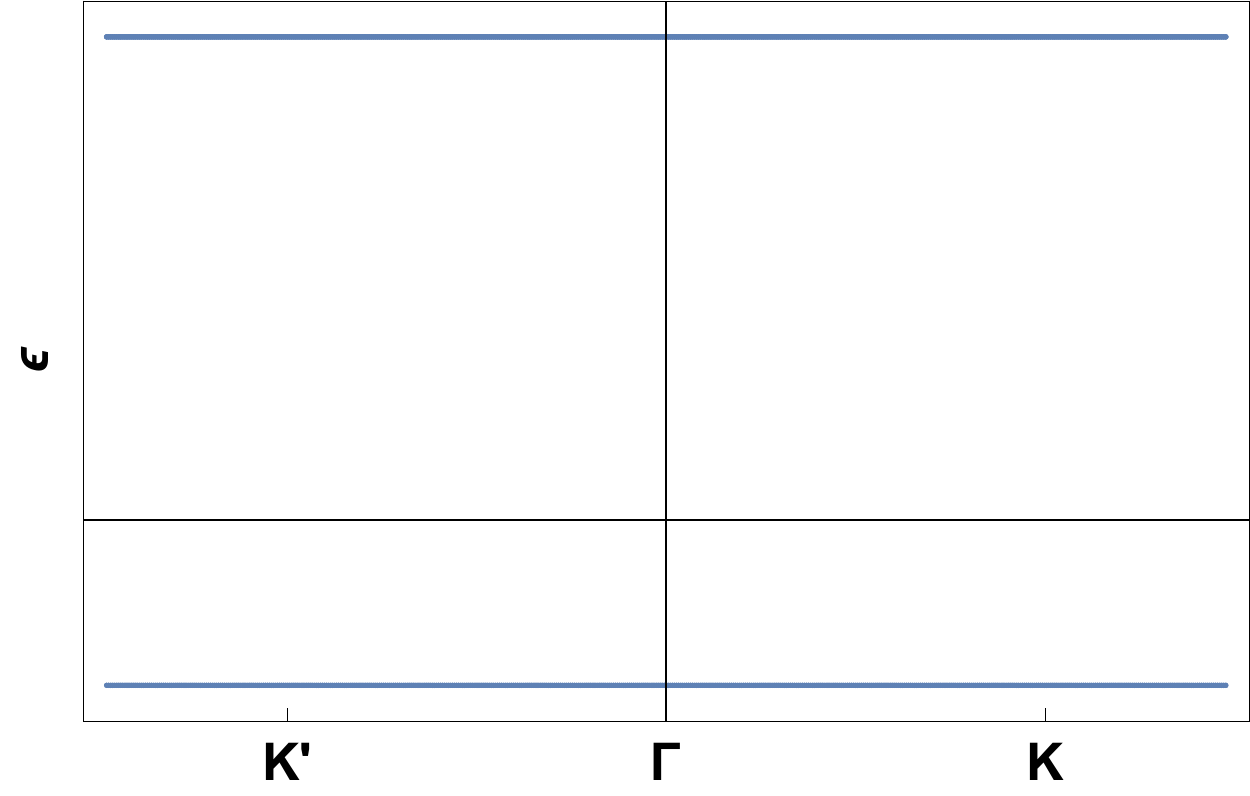}}
\caption{Bands from the network model with $\lambda =\pi/3$ and evenly spaced $\tau \in [0,\pi/2]$.}
\label{fig3_supp}
\end{figure}

We see that $\tau = 0$ corresponds to the original model, and $\tau = \pi/2$ corresponds to complete backscattering.  Figure \ref{fig3_supp} shows the resulting bands for partial backscattering.  We observe that one of the flat bands can be passed between the dispersive bands. Also, the particle-hole symmetry of the dispersive bands is lost. This is precisely the deviation we see in the tight binding numerics of the main text, suggesting that we are including much of the relevant scattering effect. 

\end{widetext}


\begin{thebibliography}{99}
\bibitem{footnote} In our model Hamiltonian  Eq.~(\ref{hamreal}) this is strictly true only at $\vartheta \neq \pi$. However, there is no nontrivial commensuration with $\vartheta = \pi$, so we do not discuss that case in this Letter.
\end{thebibliography}
\end{document}